\documentclass{article}

\usepackage{arxiv}
\usepackage[ruled,vlined]{algorithm2e}

\usepackage[utf8]{inputenc} 
\usepackage[T1]{fontenc}    
\usepackage{hyperref}       
\usepackage{url}            
\usepackage{booktabs}       
\usepackage{amsfonts}       
\usepackage{nicefrac}       
\usepackage{microtype}      
\usepackage[acronym,smallcaps]{glossaries}
\usepackage{graphicx}
\usepackage{amsmath}
\usepackage{tikz}
\usepackage{cleveref}[2012/02/15]
\usepackage{dsfont}
\usepackage{multicol}
\usepackage{bbm}
\usepackage{enumitem}
\usepackage{verbatim}
\usepackage[title]{appendix}
\usepackage{enumerate}
\usepackage{bbm}

\usepackage{lmodern}
\usepackage{amsmath}
\usepackage[ruled, vlined]{algorithm2e}

\usetikzlibrary{shapes, positioning, fit, calc, bayesnet, decorations.pathreplacing, matrix}

\newcommand*{\E}{\ensuremath{\mathbb{E}}}

\newcommand*{\yit}{\ensuremath{\mathbf{y}_{it}}}

\newcommand*{\itj}[1]{\ensuremath{#1_{it}^{(j)}}}

\newcommand*{\betafunc}[2]{\ensuremath{\text{B}\left(#1, #2\right)}}
\newcommand*{\normaldist}[2]{\ensuremath{\mathcal{N}\left(#1, #2\right)}}
\newcommand*{\deltafunc}[1]{\ensuremath{\delta\left(#1\right)}}
\newcommand*{\conditionaldist}[2]{\ensuremath{p(#1 \mid \mathbf{y}_{0:#2})}}
\newcommand*{\smoothingdist}[1]{\conditionaldist{#1}{T}}
\newcommand*{\filteringdist}[1]{\conditionaldist{#1}{t}}
\newcommand*{\predictivedist}[1]{\conditionaldist{#1}{t-1}}
\newcommand*{\ji}[1]{\ensuremath{#1_{i}^{(j)}}}
\newcommand{\indep}{\perp \!\!\! \perp}
\newcommand*{\itlast}[1]{\ensuremath{#1_{it}^{(T-t)}}}
\newcommand*{\ilast}[1]{\ensuremath{#1_{i}^{(T-t)}}}
\newcommand*{\tlast}[1]{\ensuremath{#1_{t}^{(T-t)}}}

\crefformat{footnote}{#2\footnotemark[#1]#3}
\title{Bayesian imputation of COVID-19 positive test counts for nowcasting under reporting lag}
\date{November 2020}

\author{
  Radka Jersakova\thanks{Equal contribution} \\
  The Alan Turing Institute\\
  \texttt{rjersakova@turing.ac.uk}
  \And
  James Lomax\footnotemark[1] \\
  The Alan Turing Institute\\
  \texttt{james.m@turing.ac.uk} \\
  \And
  James Hetherington \\
  Joint Biosecurity Centre, and \\
  The Alan Turing Institute \\
  \texttt{jhetherington@turing.ac.uk}
  \And
  Brieuc Lehmann \\
  University of Oxford\\
  \texttt{brieuc.lehmann@bdi.ox.ac.uk} \\
  \And
  George Nicholson \\
  University of Oxford\\
  \texttt{george.nicholson@stats.ox.ac.uk} \\
  \And
  Mark Briers \\
  The Alan Turing Institute\\
  \texttt{mbriers@turing.ac.uk} \\
  \And
  Chris Holmes \\
  University of Oxford, and\\
  The Alan Turing Institute\\
  \texttt{cholmes@turing.ac.uk}
}

\newacronym{bnn}{BNN}{Bayesian Neural Network}
\newacronym{vi}{VI}{Variational Inference}
\newacronym{hmc}{HMC}{Hamiltonian Monte Carlo}
\newacronym{mc}{MC}{Monte Carlo}
\newacronym{mcmc}{MCMC}{Markov Chain Monte Carlo}
\newacronym{fgsm}{FGSM}{Fast Gradient Sign Method}
\newacronym{sgd}{SGD}{Stochastic Gradient Descent}
\newacronym{map}{MAP}{Maximum A Posteriori}
\newacronym{postexp}{PE}{Posterior Expectation}
\newacronym{pgd}{PGD}{Projected Gradient Descent}
\newacronym{elbo}{ELBO}{Evidence Lower Bound}
\newacronym{kld}{KL}{Kullback–Leibler divergence}
\newacronym{gnj}{GNJ}{Group Normal-Jeffreys}
\newacronym{ghs}{GHS}{Group Horseshoe}
\newacronym{ppe}{PPE}{Posterior Point Estimate}
\newacronym{eot}{EOT}{Expectation Over Transformations}
\newacronym{cw}{C&W}{Carlini \& Wagner}
\newacronym{auroc}{AUROC}{Area Under ROC Curve}
\newacronym{rssi}{RSSI}{Received Signal Strength Indicator}
\newacronym{mag}{MAG}{Modelling and Analytics Group}
\newacronym{jbc}{JBC}{Joint Biosecurity Centre}
\newacronym
[
  longplural={Upper-Tier Local Authorities}
]
{utla}{UTLA}{Upper-Tier Local Authority}
\newacronym
[
  longplural={Lower-Tier Local Authorities}
]
{ltla}{LTLA}{Lower-Tier Local Authority}
\newacronym{gmrf}{GMRF}{Gaussian Markov Random Field}
\newacronym{mh}{MH}{Metropolis Hastings}
\newacronym{phe}{PHE}{Public Health England}
\newacronym{smc}{SMC}{Sequential Monte Carlo}
\newacronym{nhs}{NHS}{National Health Service}
\newacronym{dag}{DAG}{Directed Acyclic Graph}
\newacronym{mae}{MAE}{Mean Absolute Error}
\newacronym{bym}{BYM}{Besag York Mollié}
\glsdisablehyper

\begin{document}

\SetKwInput{KwInit}{Initialise}
\SetKwInput{KwSet}{Set}
\SetKwInput{KwCompute}{Compute}
\SetKwInput{KwConstruct}{Construct}
\SetKwInput{KwSample}{Sample}

\maketitle

\begin{abstract}
Obtaining up to date information on the number of UK COVID-19 regional infections is hampered by the reporting lag in positive test results for people with COVID-19 symptoms. In the UK, for ``Pillar 2'' swab tests for those showing symptoms, it can take up to five days for results to be collated. We make use of the stability of the under reporting process over time to motivate a statistical temporal model that infers the final total count given the partial count information as it arrives. We adopt a Bayesian approach that provides for subjective priors on parameters and a hierarchical structure for an underlying latent intensity process for the infection counts. This results in a smoothed time-series representation now-casting the expected number of daily counts of positive tests with uncertainty bands that can be used to aid decision making. Inference is performed using sequential Monte Carlo. 

\end{abstract}

\clearpage

\section{Introduction}

In light of COVID-19, the UK government tracks the number of lab-confirmed positive tests over time, primarily as a measure of progress against epidemic control\footnote{Tests here refer to PCR swab tests which detect presence of the virus. We analyse number of tests carried out through the \gls{nhs} and \gls{phe} as well as commercial partners (pillars 1 and 2).}. Since it takes time for test results to be reported to their local authority, and subsequently centralised, there is uncertainty on the most recent positive test counts. This uncertainty diminishes over time until all tests for a particular day are eventually reported, whereafter the count remains unchanged. The time taken until the reported counts converge to a final value, here referred to as \textit{reporting lag}, is around four days. News reports and publicly available summaries of the positive tests\footnote{\label{phe_dash}Summary statistics and visualisations of the latest available data are presently available as a dashboard at \url{https://coronavirus.data.gov.uk}.} ignore the days for which the counts have not yet converged to a final value, and often report a moving average of the positive tests.

We propose here a model on the positive test count with reporting lag which enables `now-casting' of the true count with uncertainty; in other words, by correcting for the underestimate in live reported data, we are able to suitably estimate and impute the actual positive test count and extend the seven-day moving average to the present moment. We also demonstrate how to incorporate the model into a statistical alerting system which is triggered when there is high confidence the reported positive test counts are above a threshold value. 

Given the pace of the COVID epidemic, there are a number of concurrent works \cite{lancaster, strathclyde, phe1, phe2} with similar features to our approach. These use either binomial or negative binomial models for test counts combined with spatio-temporal models (an approach widely used in epidemiology for modelling disease risk and spread). In contrast to our model, however, they do not consider reporting lag, and only analyse data once all the results are in.

\begin{figure}
    \centering
    \includegraphics[width=\textwidth]{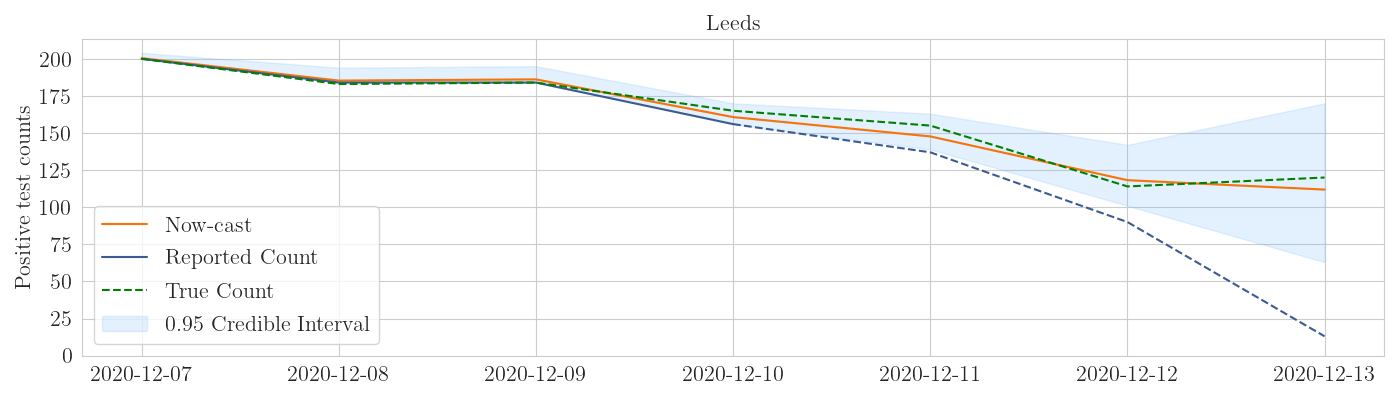}
    \caption{Example now-cast for Leeds in December.}
    \label{fig:example_timeseries}
\end{figure}

We demonstrate that the reporting lag is in fact predictable, and include it in our model to return a now-cast that incorporates the most recently available reported data. We model the reporting lag using binomial thinning; whilst there already exist well-founded mechanisms for building auto-regressive binomial thinning models on count data \cite{jung2006}, we choose instead to learn the thinning rates directly from empirical data to avoid restricting the dynamics of the lag to a particular form or order of auto-regressive prior. With this approach we gain the additional benefit of finite-time saturation to a fixed value, which is a property observed in all sequences of lagged reports. We combine empirical priors on the under-reporting and a generative model in a time series prior, providing a posterior distribution on the underlying intensity of the disease. In Figure \ref{fig:example_timeseries} we show the posterior distribution on the time series of counts for Leeds in December alongside the true and reported counts as an example of now-casting with uncertainty quantification.

Our approach is similar to two recent works that model a lagged reporting process for count data applied to COVID deaths \cite{Seaman2020.09.15.20194209} and cases of dengue fever \cite{stoner2020}; both introduce temporal structure and encode week-day and calendar effects into the reporting lag. One of the key differences between our models is the prior on the latent disease intensity. Instead of cubic splines, we impose a random walk prior with drift.

\section{Data}

The collection of daily lab-confirmed COVID-19 positive test counts are available as open data\footnote{ See footnote \ref{phe_dash}}. The data are stratified, so that the count for a nation is equal to the sum of counts for its regions. These regions are themselves divided into \glspl{utla}, and each of these \glspl{utla} is covered by a set of \glspl{ltla}, with the highest resolution count data available at the \gls{ltla}-level. In England, there are 9 regions, 150 \glspl{utla}, and 316 \glspl{ltla}. 

\begin{figure}
    \centering
    \includegraphics[width=\textwidth]{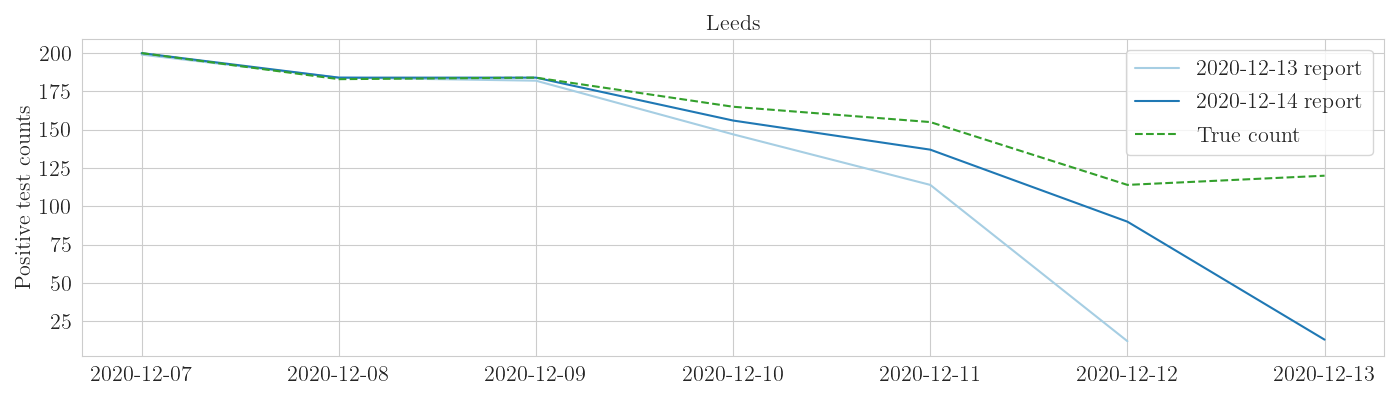}
    \caption{Two consecutive reports of positive test counts for Leeds in December. The report on 14\textsuperscript{th} December updates counts reported on 13\textsuperscript{th} for test dates from 10\textsuperscript{th} December onwards, before which they each agree with the true count. Both underestimate the true count for the most recent dates, and will therefore continue to be updated in subsequent reports until the true count is reached.}
    \label{fig:example_reports}
\end{figure}

On each day, every \gls{ltla} reports a sequence of positive test counts for all test dates up to but not including the current day, allowing for updates to previously reported values (Figure \ref{fig:example_reports}). The most recently reported counts consistently underestimate the true count due to the lag in reporting. As time progresses and more tests are processed, the reported value for a given test date approaches the true count with increasing certainty (Figure \ref{fig:example_reports}). As a result, for each test date we observe a sequence of monotone increasing reports which eventually converges to the correct count for that particular date.\footnote{In rare cases an error can lead to an over-estimate report followed by a lower count update.}

\section{Notation}
Let $i\in\{1,\ldots, 316\}$ index the collection of \glspl{ltla}. Let $t \in \{0,\ldots, T\}$ index the number of days for which data is available so that $x_{it}$ is an unobserved random variable corresponding to the true count for \gls{ltla} $i$ on day $t$. Let $j\in\{1, \ldots, T - t\}$ index the reporting lag so that $y_{it}^{(j)}$ denotes the report for day $t$ on day $t+j$. Each true count $x_{it}$ is associated with a sequence of observed, reported counts $\mathbf{y}_{it} = (y_{it}^{(1)},\ldots,y_{it}^{(T-t)})$. For some finite but unknown maximum reporting lag $\tau_{it}$, we observe  $y_{it}^{(j)} = x_{it}$ for $j > \tau_{it}$. 

Our aim is to specify a model on the true counts $x_{it}$, given reported counts $y_{it}^{(j)}$ and without knowledge of $\tau_{it}$, in order to infer a distribution on the $x_{it}$ which concentrates on the true value as $T$ increases. We further define the reporting rate at lag $j$,  $\itj{\theta} := \frac{\itj{y}}{x_{it}}$, to be the proportion of the true count that is reported at lag $j$. For historical data such that $y_{it}^{(T-t)}$ has converged to $x_{it}$, we can study $\itj{\theta}$ in order to characterise the behaviour of the reporting lag. In the following sections, we omit subscripts on $\itj{\theta}$ to indicate the random variable obtained by marginalising with respect to that subscript. For example, $\theta_i^{(j)}$ is the random variable taking value $\itj{\theta}$ with probability $1/T$. 
\section{Reporting lag} \label{sec:eda}

\begin{figure}
    \centering
    \includegraphics[width=\textwidth]{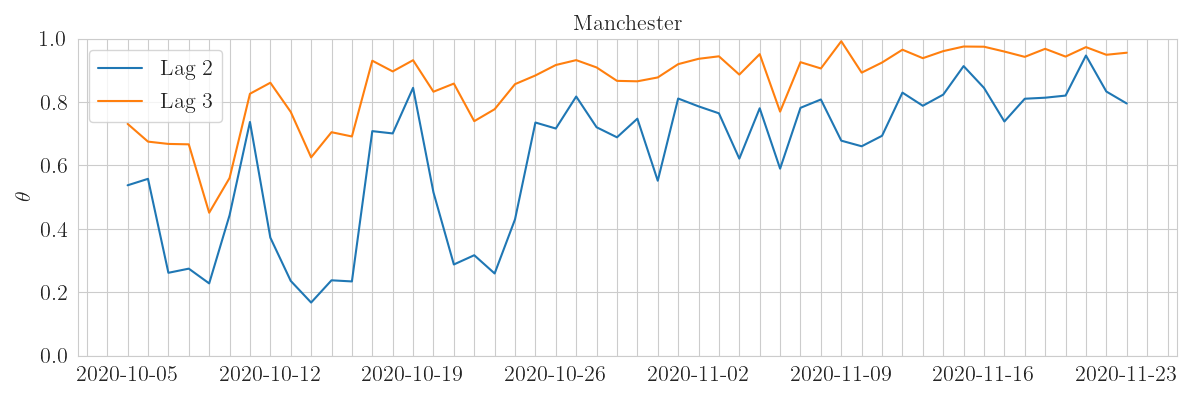}
    \caption{Time series of the true reporting rate $\theta$ at lag 2 and 3 for Manchester in October and November. The reporting rate is not temporally stable in October but becomes more predictable in November.}
    \label{fig:reporting_history}
\end{figure}

\begin{figure}
    \centering
    \includegraphics[width=\textwidth]{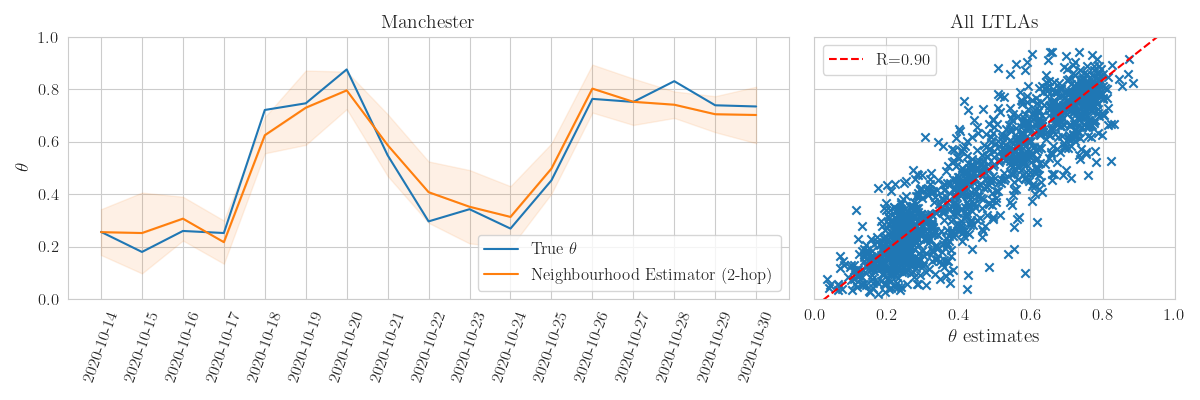}
    \caption{Lag-2 spatial neighbourhood priors. \emph{Left}: True reporting rate $\theta^{(2)}_{it}$ for Manchester, and an estimator built from its 2-hop \gls{ltla} neighbours for each day from the 14\textsuperscript{th} to the 30\textsuperscript{th} of October and \emph{Right}: 2-hop \gls{ltla} neighbour estimator against the true $\theta^{(2)}_{it}$ across all \glspl{ltla} with at least one reported count $x_{it} > 100$, combined over all days between the 14\textsuperscript{th} to the 30\textsuperscript{th} of October.}
    \label{fig:spatial_lag_2}
\end{figure}

We develop our now-casting model on the basis of the empirical behaviour of the reporting rates $\itj{\theta}$, which we now describe. Since July 2020, we have observed two distinct modes of predictable reporting behaviour which we refer to as spatially and temporally stable. 

\subsection{Temporally stable reporting} \label{sec:temporally_stable}
When reporting is temporally stable, we observe that the reporting rates for an \gls{ltla} at each lag do not change much with time and may therefore be estimated by

\begin{align}
\label{eq:theta_est_temporal}
    \hat{\theta}_{it}^{(j)} = \frac{1}{T-1}\sum_{k\neq t}^{T}\theta_{ik}^{(j)}
\end{align}

\begin{figure}
    \centering
    \includegraphics[width=\textwidth]{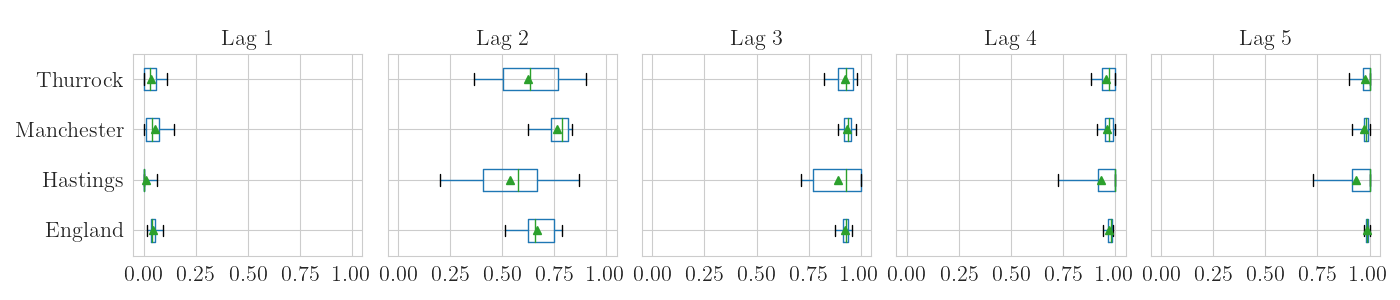}
    \caption{Empirical distributions of the reporting rates $\ji{\theta}$ in November for a selection of LTLAs. We also include England to show that the marginal $\theta^{(j)}$ obeys the intuition encoded by observations (i) and (ii) in Section \ref{sec:eda}; the mean of $\ji{\theta}$ increases with increasing lag, indicating that more reports are accounted for as time progresses.}
    \label{fig:all_ltla_theta}
\end{figure}

Figure \ref{fig:all_ltla_theta} shows empirical distributions for $\itj{\theta}$ marginally across the day-index $t$ for lags $j=\{1, \ldots, 5\}$. We observe that $\E[\theta^{(j)}]$ is increasing in $j$.

The reporting rates are predictable in a fashion supported by our intuition; as lag increases we are increasingly confident that $\itj{\theta}=1$ across all \glspl{ltla}, and this state is reached by a coherent sequence of updates to an initial underestimate. It is also clear from Figure \ref{fig:all_ltla_theta} that there is enough variation in $\ji{\theta}$ between \glspl{ltla} to warrant their separate consideration for modelling. When reporting behaviour changes slowly with time, we may construct a temporally local set  $\mathcal{S}_{it}^{(j)} := \{t^\prime: t - W \leq t^\prime \leq t - \tau_{it^\prime}\}$ where $W$ is the length of the stable interval, and update equation \ref{eq:theta_est_temporal} so that $\theta_{it}^{(j)}$ is estimated by

\begin{align}
\label{eq:theta_est_temporal}
    \hat{\theta}_{it}^{(j)} = \frac{1}{T-1}\sum_{k\in \mathcal{S}_{it}^{(j)}}\theta_{ik}^{(j)}
\end{align}

\subsection{Spatially stable reporting} \label{sec:spatially_stable}

Let $N_n\left(i\right)$ be the $n$-hop neighbourhood of \gls{ltla} $i$ on the adjacency graph of \glspl{ltla}. When reporting is spatially stable we observe that the reporting rates of \glspl{ltla} which are close to one-another are similar and so we may estimate a reporting rate for an \gls{ltla} from those of its neighbours by

\begin{align} \label{eq:spatial_estimator}
    \hat{\theta}_{it}^{(j)} = \frac{1}{|N_n\left(i\right)|} \sum_{k\in N_n\left(i\right)} \theta_{kt}^{(j)}
\end{align}

In the left panel of Figure \ref{fig:spatial_lag_2} we show the performance of a 2-hop neighbourhood estimator for Manchester in October; the reporting rates of neighbouring \glspl{ltla} track one another. It is clear though that the reporting is not temporally stable, and so we must rely on spatial estimates alone. In the right panel we measure the performance of the 2-hop estimates \eqref{eq:spatial_estimator} against the truth for all \glspl{ltla} where we have observed at least one report $x_{it} > 100$ marginally across the dates between the 14\textsuperscript{th} and 30\textsuperscript{th} October. There is a clear linear relationship $(\text{R}=0.9)$ with the truth.

\subsection{Empirical Bayes prior specification} \label{sec:prior_estimation}

\begin{figure}
    \centering
    \includegraphics[width=\textwidth]{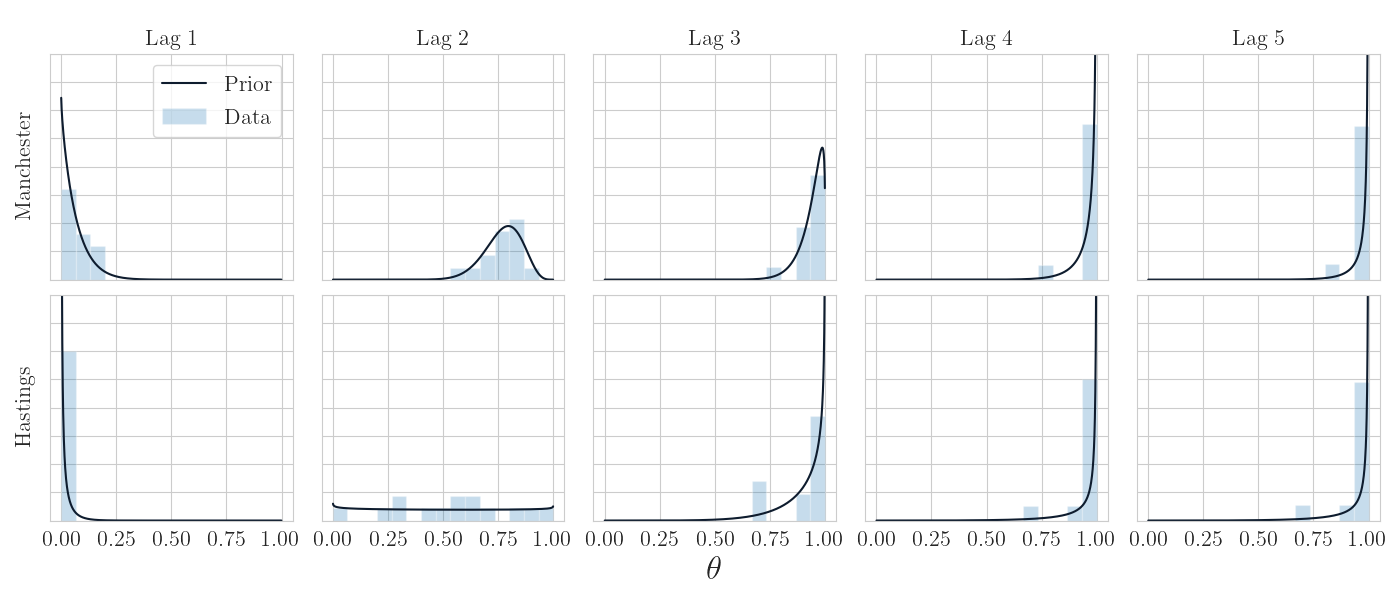}
    \caption{Empirical distributions and moment matched Beta priors $\ji{\theta}$ for Manchester and Hastings in November.}
    \label{fig:example_theta_priors}
\end{figure}

In each of sections \ref{sec:temporally_stable} and \ref{sec:spatially_stable} we demonstrate that noisy estimates of the reporting rates $\theta_{it}^{(j)}$ can be constructed. In order to avoid underdispersion in models on $x_{it}$ we must capture uncertainty in these estimates. We therefore propose that $\itj{\theta} \sim \text{Beta}(\ji{\alpha}, \ji{\beta})$ and estimate $\ji{\alpha}$ and $\ji{\beta}$ by moment matching with the empirical distribution measured on the most recent 2 weeks of converged reports. Denote by $m_{ij}=\E[\ji{\theta}]$ and $v_{ij}=\text{Var}[\ji{\theta}]$ the means and variances of these empirical distributions and let $\nu_{ij} = \text{min}\{v_{ij}, m_{ij}(1 - m_{ij}) - \epsilon\}$ with $\epsilon > 0$ but arbitrarily small, then

\begin{align}
    \ji{\alpha} = \frac{m_{ij}^2 \left(1 - m_{ij}\right)}{\nu_{ij}} - m_{ij} \qquad \qquad \ji{\beta} = \ji{\alpha} \left(\frac{1 - m_{ij}}{m_{ij}}\right) \label{eq:ab_moment}
\end{align}

are chosen. Figure \ref{fig:example_theta_priors} illustrates the empirical distribution and resulting fit for two \glspl{ltla} in November - as time progresses, the moment matching produces Beta distributions with most of their mass on large values of $\theta_{it}^{(j)}$ as expected.

\section{Modelling} \label{sec:modelling}

Based on the empirical observations made in Section \ref{sec:eda}, we now describe a model on reports $y^{(j)}_{it}$ which accounts for a predictable reporting lag in order to impute up-to-date predictive distributions on the current true count $x_{it}$. We model the reported counts $y^{(j)}_{it}$ as a random variable following a binomial distribution conditional on the (unobserved) true count $x_{it}$ and the reporting rate $\theta_{it}^{(j)}$, placing a beta prior on $\theta_i^{(j)}$ with parameters $\alpha_i^{(j)}, \beta_i^{(j)}$ determined by \eqref{eq:ab_moment}. Modern research on epidemic monitoring typically poses negative binomial models on count data, with spatio-temporal regularization for smoothness; here we capture the same effect by proposing that the $x_{it}$ result from a Poisson process and integrating over the rate parameter under a Gaussian random-walk time-series prior. We introduce the model in a modular fashion.

\subsection{Binomial thinning}
\label{sec:binomial_thinning}

To describe the relationship between the true counts $x_{it}$ and the reported counts $\yit$, we treat each $y_{it}^{(j)}$ as a binomial random variable with success probability $\itj{\theta}$ over $x_{it}$ independent Bernoulli trials. Following the arguments of Section \ref{sec:eda}, we place beta priors on the $\itj{\theta}$ with moment-matched $(\ji{\alpha}, \ji{\beta})$. The most recent available report $y_{it}^{(T-t)}$ is a sufficient statistic for $x_{it}$  (Appendix \ref{appendix:suff_stat_x}) and so we build the following hierarchical model:

\begin{align}
  \itlast{\theta} &\sim \text{Beta}\left(\ilast{\alpha}, \ilast{\beta}\right)\\
  \itlast{y} \mid \itlast{\theta}, x_{it} &\sim \text{Binomial}\left(\itlast{\theta}, x_{it}\right)
  \label{eq:y_binomial}
\end{align}

Integrating out each $\itlast{\theta}$ yields the following joint distribution on $(x_{it}, \itlast{y})$:

\begin{align}
  p(x_{it}, \itlast{y}) &= p(x_{it})p(\itlast{y} |x_{it}) \\ 
  &= p(x_{it}){x_{it}\choose{\itlast{y}}}\frac{\betafunc{\itlast{y} + \ilast{\alpha}}{x_{it} - \itlast{y} + \ilast{\beta}}}{\betafunc{\ilast{\alpha}}{\ilast{\beta}}}
  \label{eq:indep_count_joint}
\end{align}

where $p(x_{it})$ is a prior on $x_{it}$ which may, for example, be flat across a set of feasible positive test counts given the population of the \gls{ltla} under consideration. This joint distribution may be used to draw samples from the marginal posterior $p(x_{it} \mid \mathbf{y}_{it})$ with \gls{mh}. 

\subsection{Latent intensity process}
\label{sec:latent_process}

Whilst estimating the true count directly is clearly important in monitoring the epidemic, a more compelling epidemiological variable is the latent rate which controls emission of these counts. To extend the model, we therefore assume that each $x_{it}$ is the result of a Poisson process with rate $\lambda_{it}$ so that the hierarchy is now given by

\begin{align}
    \lambda_{it} &\sim \text{Gamma}\left(a_{it}, b_{it}\right) \label{eq:gam_prior}\\
    \itlast{\theta} &\sim \text{Beta}\left(\ilast{\alpha}, \ilast{\beta}\right)\\
    x_{it} \mid \lambda_{it} &\sim \text{Poisson}\left(\lambda_{it}\right)\\
    \itlast{y} \mid \itlast{\theta}, x_{it} &\sim \text{Binomial}\left(\itlast{\theta}, x_{it}\right)
\end{align}

When $\lambda_{it}$ is integrated out under a gamma prior distribution this is equivalent to proposing a negative-binomial distribution on $x_{it}$, which is a common choice in modelling count data for epidemic monitoring. The joint distribution on $(\lambda_{it}, \itlast{y})$ is given by:

\begin{align}
  p(\lambda_{it}, \itlast{y}) = p(\lambda_{it})\sum_{x_{it}}\frac{\lambda^{x_{it}} e^{-\lambda_{it}}}{x_{it}!} {x_{it}\choose{\itlast{y}}}\frac{\betafunc{\itlast{y} + \ilast{\alpha}}{x_{it} - \itlast{y} + \ilast{\beta}}}{\betafunc{\ilast{\alpha}}{\ilast{\beta}}} \label{eq:poisson_beta_binom}
\end{align}

which may be used as a \gls{mh} potential function in order to generate samples from the new posterior of interest $p(\lambda_{it} \mid \itlast{y})$.

\subsection{Temporal smoothing}\label{sec:time_dependent_models}

Estimates of the latent rates $\lambda_{it}$ may suffer from high variance, particularly in the early stages of reporting when the reported $\itlast{y}$ likely constitute underestimates of $x_{it}$. To reduce this variance, we encode time dependence in the latent rates, as follows. Let $\kappa_{it}$ be the difference between Poisson rates so that

\begin{align} \label{eq:rw_prior}
  \lambda_{it} = \kappa_{it} + \lambda_{i,t-1}
\end{align}

Further impose an AR1 prior with scale $\sigma_i$ on the sequence $\kappa_{i,0:T}$ so that given a standard normal random variable $\epsilon_t \sim \normaldist{0}{1}$ we may write

\begin{align} \label{eq:rw_prior_kappa}
    \kappa_{it} = \kappa_{i,t-1} + \sigma_i\epsilon_t \qquad \Longleftrightarrow \qquad p(\kappa_{it} | \kappa_{i,t-1}) = \normaldist{\kappa_{i,t-1}}{\sigma_i^2}.
\end{align}

This represents an intuitive assumption of local temporal smoothness in the epidemic rate and is a common choice of temporal regularisation; the Kalman Filter \cite{kalmanfilter}, to which our model bears close resemblance, operates under a similar latent transition distribution. The prior induces dependence between each $\lambda_{it}, \kappa_{it}$ and the observed data $\mathbf{y}_{i, 0:T} = \mathbf{y}_{i,0}, \ldots ,\mathbf{y}_{i,T}$. The key distributions of interest are then the joint \emph{filtering} and \emph{smoothing} distributions given by $p(\lambda_{it}, \kappa_{it} \mid \mathbf{y}_{i, 0:t})$ and $p(\lambda_{it}, \kappa_{it} \mid \mathbf{y}_{i, 0:T})$ respectively.

\subsection{Weekend effects} \label{sec:weekend_effect_model}

\begin{figure}
    \centering
    \includegraphics[width=\textwidth]{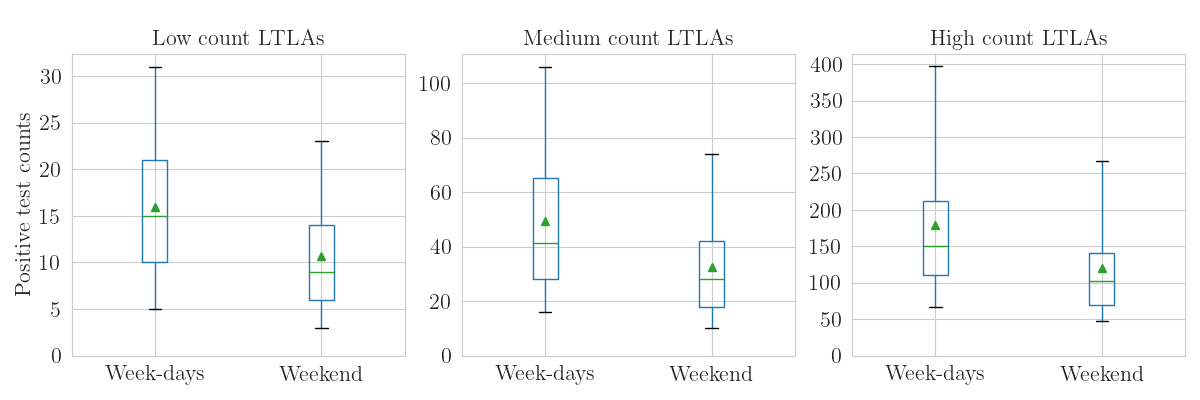}
    \caption{Daily positive test counts reported for week-day and weekend test dates in November across all LTLAs. The LTLAs are classified based on whether the mean number of tests is $<20$ (Low count LTLAs), $<100$ (Medium count LTLAs) or $>=100$ (High count LTLAs). We observe some reduction in tests on weekends relative to week-days across all LTLAs.}
    \label{fig:weekend_test_effects}
\end{figure}

The smoothness assumption described in Section \ref{sec:time_dependent_models} constrains the sequence $\lambda_{i,0:T}$ to vary in proportion to the random walk scale $\sigma_{i}$. The weekend effects demonstrated in Figure \ref{fig:weekend_test_effects} break this smoothness assumption; when there are predictable drops in the test count at weekends it is because fewer tests are taken, rather than any true decrease in the underlying incidence rate of the disease. To capture this effect, we introduce latent random variables $z_{i,t} \in [0,1]$ with prior distribution

\begin{align}
    p(z_{i,t}) = \begin{cases}
      \delta\left(z_{i,t} - 1\right) & \text{if $day(t) \in \{\text{Monday}, \ldots, \text{Friday}\}$ }\\
      \text{Beta}\left(a,b\right) & \text{if $day(t) \in \{\text{Saturday},\text{Sunday}\}$}
    \end{cases}
\end{align}

then let the emission distribution on $x_{it}$ be

\begin{align}
    x_{it} \mid \lambda_{it}, z_{it} \sim \text{Poisson}\left(z_{it} \lambda_{it}\right)
\end{align}

so that smoothness in $\lambda_{i,0:T}$ is maintained by allowing $z_{it}$ to capture these observed weekend effects by emitting counts at a reduced rate. In practice a flat $\text{Beta}\left(1,1\right)$ prior allows the smoothness assumption to be maintained, though selecting a stronger $a$ and $b$ may be possible if weekend effects are predictable in their magnitude. We can measure the strength of these effects by examining the posterior smoothing distributions $p(z_{it} \mid \mathbf{y}_{0:T})$ on weekend days. In Appendix \ref{sec:appendix_smoothing} we give details on how to evaluate this posterior in the case where each weekend day has its own unique latent effect, but share prior parameters, as well as demonstrating how this addition alters the procedure for determining $p(x_{it} \mid \mathbf{y}_{i, 0:T})$. From now on we omit the \gls{ltla} index $i$ for brevity.

\subsection{The complete model} \label{sec:complete_model}

\begin{figure}
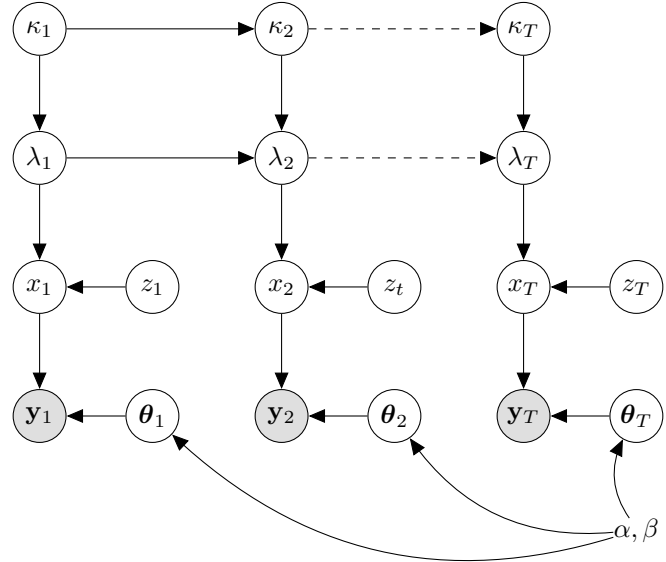
 
    \centering
    
        \tikz{ %
            \node[obs] (y) {$\mathbf{y}_1$} ; %
            \node[latent, above=of y] (x) {$x_1$} ; %
            \node[latent, above=of x] (lam) {$\lambda_1$} ; %
            \node[latent, above=of lam] (kap) {$\kappa_1$} ; %
            \node[latent, above=of y, xshift=1.5cm] (z) {$z_1$};
            \node[latent, below=of z] (t1) {$\boldsymbol{\theta}_1$};
            
            \node[latent, right=of z] (xt) {$x_2$} ;
            \node[obs, below=of xt] (yt) {$\mathbf{y}_2$} ;
            \node[latent, above=of xt] (lamt) {$\lambda_2$} ;
            \node[latent, above=of lamt] (kapt) {$\kappa_2$} ; %
            \node[latent, above=of yt, xshift=1.5cm] (zt) {$z_t$};
            \node[latent, below=of zt] (t2) {$\boldsymbol{\theta}_2$};
            
            \node[latent, right=of zt] (xT) {$x_T$} ;
            \node[latent, above=of xT] (lamT) {$\lambda_T$} ;
            \node[latent, above=of lamT] (kapT) {$\kappa_T$} ; %
            \node[obs, below=of xT] (yT) {$\mathbf{y}_T$} ;
            \node[latent, above=of yT, xshift=1.5cm] (zT) {$z_T$};
            \node[latent, below=of zT] (t3) {$\boldsymbol{\theta}_T$};

            \node[const, below=of t3    ] (theta) {$\alpha, \beta$} ;

            \edge {x}{y} ; %
            \edge {xt}{yt} ;
            \edge {xT}{yT} ;
            
            \edge {kap}{lam} ;
            \edge {kapt}{lamt}
            \edge {kapT}{lamT} ;

            \edge {lam}{x} ;
            \edge {lamt}{xt}
            \edge {lamT}{xT} ;

            \edge {lam}{lamt} ;
            \edge[dashed] {lamt}{lamT} ;
            
            \edge {kap}{kapt} ;
            \edge[dashed] {kapt}{kapT} ;
            
            \edge {z}{x} ;
            \edge {zt}{xt} ;
            \edge {zT}{xT} ;
            
            \edge {t1}{y}
            \edge {t2}{yt}
            \edge {t3}{yT}
            
            \path (theta) edge[bend left, ->] (t1) ; %
            \path (theta) edge[bend left, ->] (t2) ; %
            \path (theta) edge[bend left, ->] (t3) ; %
            
          }
    \caption{\gls{dag} depicting the dependencies in the model with joint distribution \eqref{eq:complete_joint}. The latent Poisson rates $\lambda_t$ are coupled by the AR1 drift $\kappa_t$ \eqref{eq:rw_prior}. Each $x_t$ arises from a Poisson process with rate $\lambda_t z_t$, where $z_t=1$ on weekdays. The report counts follow a beta-binomial distribution with parameters $x_t, y_t, \alpha, \beta$ which results from integrating out each $\theta_t^{(j)}$ under a $\text{Beta}(\alpha^{(j)}, \beta^{(j)})$ prior. }
    \label{fig:tikz_model}
\end{figure}

Together Sections \ref{sec:binomial_thinning}-\ref{sec:weekend_effect_model} specify a smooth time-dependent model for count data exhibiting weekend effects and a lagged reporting process. The \gls{dag} in Figure \ref{fig:tikz_model} shows the full conditional dependency structure for this model, encoded equivalently by the joint distribution

\begin{align} \label{eq:complete_joint}
     p(\lambda_0) \prod_{t=0}^T \left[p(x_t \mid \lambda_t, z_t)p(\lambda_{t+1} \mid \lambda_t, \kappa_{t+1})p(\kappa_{t+1} \mid \kappa_t)p(z_t)p(y_t^{(T-t)} \mid x_t, \theta_t^{(j)})p(\theta_t^{(j)})\right].
\end{align}

We can derive conditional independence relations between variables at different time points as follows. For day $t$, denote by $\boldsymbol{\theta}_t = \{\theta_t^{(j)}\}_{j=t+1}^T$ the collection of under-reporting rates for each lagged report and by $\Omega_t = \{\lambda_t, \kappa_t, x_t, z_t, \boldsymbol{\theta}_t, \mathbf{y}_t\}$ the collection of latent variables and observations, then applying the d-separation criteria \cite{pearl2009} to the \gls{dag} in Figure \ref{fig:tikz_model} we have the conditional independence relation

\begin{align} \label{eq:conditional_indep}
    \Omega_t \indep \Omega_{\neq t} \mid \lambda_t, \kappa_t.
\end{align}

We will use these conditional independence relations as the basis for drawing samples sequentially from the posterior. 

\section{Posterior inference} \label{sec:inference}

\subsection{Metropolis Hastings Samplers} \label{sec:basic_mh}
For each of the submodels discussed in Section \ref{sec:modelling} we draw posterior samples with standard \gls{mcmc} methods. For the time-independent submodels, equations \eqref{eq:indep_count_joint} and \eqref{eq:poisson_beta_binom} serve as potential functions for simple \gls{mh} sampling. In equation \eqref{eq:indep_count_joint} the sum over $x_{t}$ can be performed directly for benign\footnote{No prior could, for example, place any mass on counts in excess of the number of tests taken in its \gls{ltla} on the day under consideration.} choices of prior on $x_{t}$, when the true $x_{t}$ is expected to be small, or by numerical integration when the prior is concentrated on a region of its support. To sample from $p(\lambda_{t}\mid \tlast{y})$ we use standard normal proposals and initialise the sampler to the mean of the prior $\E_{p(\lambda_{t})}[\lambda_{t}]$. Since the expected distance for an n-step 1-dimensional Gaussian random walk is $\sqrt{n}$, we can be confident that the posterior is well explored by choosing an $n$ such that our worst-case estimate of the absolute error $|x_{t} - \E_{p(\lambda_{t})}[\lambda_{t}]|$ is well explored by $\sqrt{n}$. In all cases we apply thinning and burn in to the sample chain, though no hard convergence checks are made.

\subsection{Marginalized particle filtering and smoothing}

\begin{figure}
    \centering
    \includegraphics[width=\textwidth]{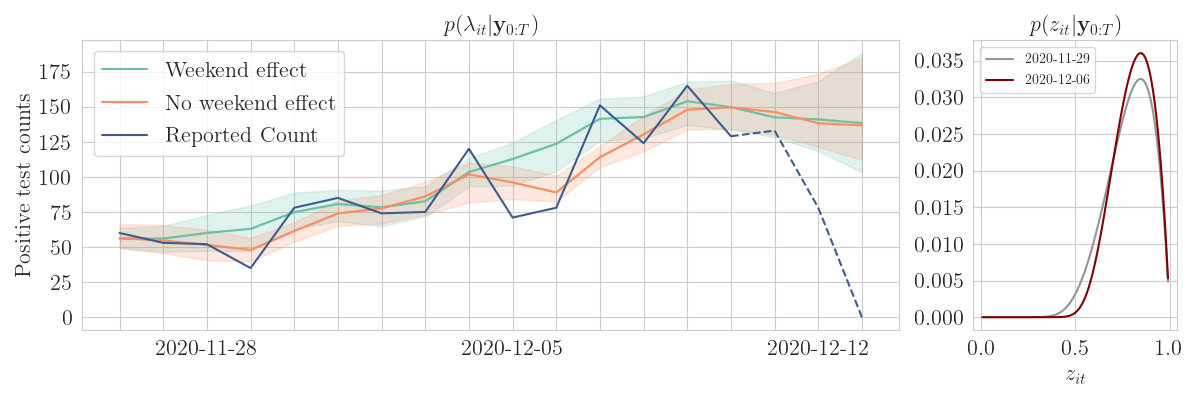}
    \caption{Smoothing distributions for $\lambda_t$ and $z_t$ in December for Canterbury with and without weekend effects. Inclusion of weekend effects allows $\lambda_t$ to remain smooth over periodic decreases in the positive test count at weekends. Each $\smoothingdist{z_{t}}$ measures the strength of the weekend effect for day $t$; on the right we show these posterior distributions for Sunday $29^{th}$ November and $6^{th}$ December.}
    \label{fig:manchester_smoothing_weekend}
\end{figure}

In the time-dependent case, we make use of the conditional independence relation \eqref{eq:conditional_indep} to construct an algorithm for filtering and smoothing which builds on the simple \gls{mh} sampling schemes as those used for inference in time-independent models. Inference mechanisms under the conditional dependencies induced by the random-walk prior in Section \ref{sec:time_dependent_models} are well studied \cite{doucet-johansen-tutorial} and so we give here the equations necessary for sequential posterior sampling of the latent $\lambda_{t}, \kappa_{t}$ and $x_{t}$. At its foundation the complete model of Section \ref{sec:complete_model} is Markov on a continuous latent state space with a non-conjugate emission distribution. We therefore employ a forward-backward algorithm, and deal with the non-conjugate structure by performing \gls{mh} sampling at each time step. The atomic distributions constructed from these samples are propagated forwards as a prior to the subsequent time step, and re-sampled backwards for smoothing; this simple particle \gls{smc} algorithm differs from the canonical bootstrap filter since each filtering distribution is sampled directly rather than re-sampled from the previous predictive distribution\footnote{We also implemented a standard bootstrap particle filter for the forward pass, but found that the accuracy of the posterior approximation worsened when the filtering predictive had poor coverage, due to high degeneracy in re-sampling particles in the tails}. 

In what follows we dispense with the \gls{ltla} index for brevity and outline the strategy for sequential sampling without weekend effects. The interested reader may refer to Appendices \ref{sec:appendix_filtering} and \ref{sec:appendix_smoothing} for derivations of marginal filtering and smoothing distributions for all latent variables and including all effects proposed in Section \ref{sec:modelling}. Consider that with samples $\Lambda_{t-1}^{(i)},K_{t-1}^{(i)} \sim \predictivedist{\lambda_{t-1}, \kappa_{t-1}}$ from the joint filtering distribution at time $t-1$ we may evaluate

\begin{align}\label{eq:filtering_joint}
  \predictivedist{\lambda_t, \kappa_t, \mathbf{y}_t} \approx \frac{p(\mathbf{y}_t \mid \lambda_t)}{N} \sum_{i=1}^N p(\lambda_t \mid \kappa_t, \Lambda_{t-1}^{(i)})p(\kappa_t \mid K_{t-1}^{(i)}) 
\end{align}

where the sum over the atoms $\Lambda_t^{(i)}, K_{t-1}^{(i)}$ results from approximating $p(\lambda_{t-1}, \kappa_{t-1} \mid \mathbf{y}_{0:t-1})$ by the atomic distribution on \gls{mh} samples. Sampling from $p(\lambda_0 \mid \mathbf{y}_0)$ is done exactly as in the time independent case using a potential given by \eqref{eq:poisson_beta_binom}; by induction we can therefore compute a sequence of joint distributions of the form \eqref{eq:filtering_joint} such that we may draw samples from the sequence of joint filtering distributions $p(\lambda_0, \kappa_0 \mid \mathbf{y}_0),\ldots ,p(\lambda_t, \kappa_t \mid \mathbf{y}_{0:t})$. We can construct the smoothing distributions by re-sampling these filtering atoms in a backward pass. Write the $j^{th}$ smoothing atoms as $L_{t+1}^{(j)}, \zeta_{t+1}^{(j)} \sim p(\lambda_{t+1}, \kappa_{t+1} \mid \mathbf{y}_{0:T})$. Then the re-sampling probabilities are

\begin{align}
  \smoothingdist{\lambda_t = \Lambda_{t}^{(i)}, \kappa_t = K_t^{(i)}} \approx \frac{1}{M}\sum_{j=1}^M \frac{w_{ij}}{w_{*j}}\label{eq:smoothing_weights}
\end{align}

where the weights and normaliser are given by $w_{ij} = \deltafunc{L_{t+1}^{(j)} - (\zeta_{t+1}^{(j)} + \Lambda_t^{(i)})}\normaldist{\zeta_{t+1}^{(j)}; K_t^{(i)}}{\sigma^2}$ and $w_{*j}=\sum_{i=1}^Nw_{ij}$ respectively. The full procedure for inference is given by Algorithm \ref{alg:inference}. In Figure \ref{fig:manchester_smoothing_weekend} we show the result of learning this smoothing distribution with and without weekend effects included. Given the smoothing atoms for each $\lambda_t$ we can compute an approximate smoothing distribution for each $x_t$ by

\begin{align} \label{eq:smooth_resample}
p(x_t \mid \mathbf{y}_{0:T}) \approx \frac{p(\mathbf{y}_t \mid x_t)}{M} \sum_{j=1}^M \frac{p(x_t \mid \zeta_{t}^{(j)})}{p(\mathbf{y}_t \mid \zeta_t^{(j)})}    
\end{align}

which for time $T$ gives us our required now-cast of the true count in the face of reporting lag.  See Figure \ref{fig:nowcast} for example now-cast showing uncertainty existing on recent counts and diminishing with time.

\begin{algorithm} \label{alg:inference}
\KwSample{N atoms $\{\Lambda_0^{(i)}\}_{i=1}^N$ from $p(\lambda_0 \mid \mathbf{y}_0)$}
\For{$t\gets1$ \KwTo T}{
    $f(\lambda_t) = p(\mathbf{y}_t \mid \lambda_t) \sum_{i=1}^N p(\lambda_t \mid \Lambda_{t-1}^{(i)})$\;
    \KwSample{N atoms $\{\Lambda_t^{(i)}\}_{i=1}^N$ by \gls{mh} on $f(\lambda_t)$}
}
\vspace{1em}
\KwSet{$\zeta_T^{(0)}, \dots \zeta_T^{(N)} \gets \Lambda_T^{(0)}, \ldots, \Lambda_T^{(N)}$}
\vspace{1em}
\For{$t\gets T-1$ \KwTo 0}{
    \KwSample{N atoms $\{\zeta_t^{(i)}\}_{i=1}^N$ by re-sampling $\{\Lambda_t^{(i)}\}_{i=1}^N$ with probabilities \eqref{eq:smoothing_weights}}
}

\Return $p(\lambda_{t} \mid \mathbf{y}_{0:T}) \approx \frac{1}{N}\sum_{i=1}^N \deltafunc{\lambda_t - \zeta_t^{(i)}} \qquad \forall t \in {0,\ldots T}$ 

\caption{\textbf{Particle \gls{smc} for smoothing distributions}}
\end{algorithm}

\subsection{Kalman Gain and Moving Averages} \label{sec:kalman_filter}

In public reporting of the positive test count, the UK government uses a windowed average computed over a seven-day period. This simple mechanism is to an extent well motivated as an approximation to linear dynamical models.  For simplicity, consider modelling only the latest reported counts for times where $y_t^{(T-t)} = x_t$  by a Kalman Filter with no drift so that

\begin{align}
    p(x_0) &= \normaldist{\mu}{\sigma^2}\\
    p(x_t \mid x_{t-1}) &= \normaldist{x_{t-1}}{\sigma^2}\\
    p(y_t^{(T-t)} \mid x_t) &= \normaldist{x_t}{\sigma_y^2}
\end{align}

Let the Kalman Gain (see Appendix \ref{sec:kalman_gain}) for time-step $t$ be $K_t \in [0,1]$. The expectation of the Kalman filtering distribution at time-step $t$ may be written as

\begin{align}
    \E_{p(x_t \mid y_{0:t})}\left[x_t\right] = K_t y_t + (1 - K_t)\E_{p(x_{t-1} \mid y_{0:t-1})}\left[x_{t-1}\right]
\end{align}

which expands recursively to give

\begin{align}
    \E\left[x_t\right] &= \mu \prod_{i=1}^t (1 - K_i) + \sum_{i=1}^t K_i y_i \prod_{j=i}^t (1 - K_j)\label{eq:kalman_weighted_sum}\\
    &= const + \sum_{i=1}^t y_i \text{w}_i
\end{align}

where $\text{w}_i = K_i \prod_{j=i}^t (1 - K_j)$. Equation \eqref{eq:kalman_weighted_sum} is a weighted average of the observed data. It is clear that $\text{w}_t > \text{w}_{\leq t - 2}$ and further that if $K_t > 1/2$ then $\text{w}_t > \text{w}_{\leq t - 1}$ so that the most recent observation has the largest weight. This gives a rough interpretation of the filtering distribution as a weighted average of the data with most weight given to the most recent observations.

\begin{figure}
    \centering
    \includegraphics[width=\textwidth]{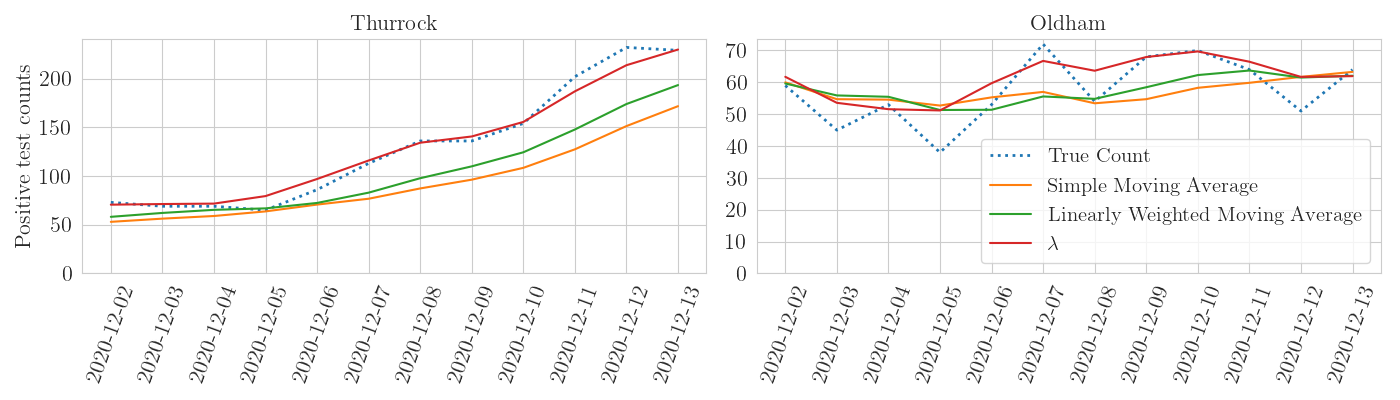}
    \caption{Smoothed counts for Thurrock and Oldham using seven day moving averages. The simple moving average weights all observations equally whereas the linearly weighted moving average gives more weight to recent observations. These are plotted alongside the expected value of the $\lambda$ smoothing posterior distribution.}
     \label{fig:moving_averages}
\end{figure}

In the absence of implementing a full model, we therefore suggest that a windowed average with decaying weights constitutes a type of posterior point-estimate for the Kalman Filter, which is in itself a reasonable approximation to our model for times where reported counts have converged to the truth; Figure \ref{fig:moving_averages} shows a comparison between our model, and a number of weighted moving averages which track the expected value of the smoothing posterior on $\lambda_t$. 

\subsection{Model selection} \label{sec:model_selection}

\begin{figure}
    \centering
    \includegraphics[width=\textwidth]{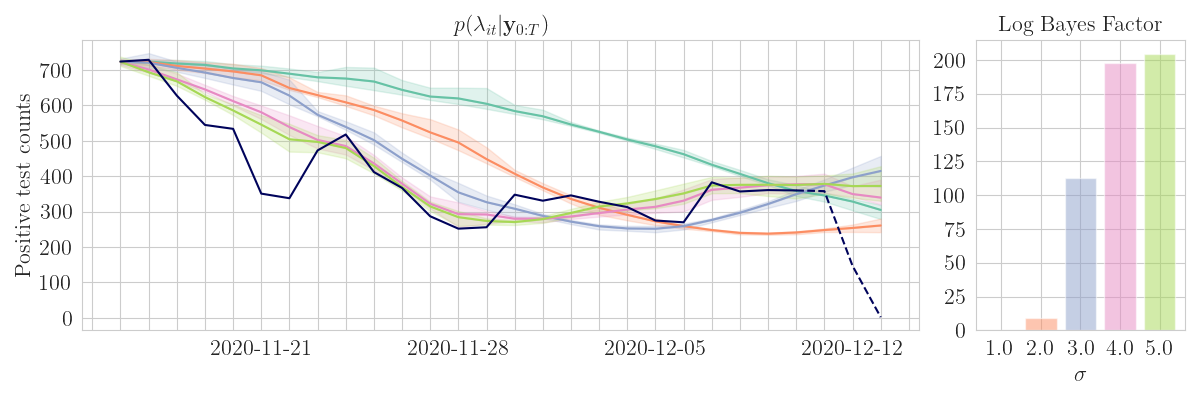}
    \caption{Smoothing distributions on the rates $\lambda_{it}$ for Birmingham and their corresponding (relative) evidence. Of those considered, the random-walk scale which maximises the evidence is $\sigma=5$.}
    \label{fig:evidence_compare}
\end{figure}

Our model includes one free parameter: the scale $\sigma$ of temporal smoothing applied by the random walk prior on the sequence $\kappa_{0:T}$ (see equation \ref{eq:rw_prior}). The choice of $\sigma$ influences the set of feasible posterior distributions on $\lambda_t$; in the absence of prior information on the rate of change of the disease it is important to choose $\sigma$ to best explain the data observed. We take an empirical Bayes approach, and maximise the model evidence\cite{fong2020}

\begin{align}
    p(\mathbf{y}_{1:T})\approx p(\mathbf{y}_0)\prod_{t=1}^{T} \frac{1}{N}\int p(\mathbf{y}_t \mid \lambda_t) \sum_{i=1}^{N} \deltafunc{\lambda_t - (\Lambda_{t-1}^{(i)} + \kappa_t)} p(\kappa_t \mid K_{t-1}^{(i)}) d\lambda_{t}d\kappa_t \label{eq:evidence_body}
\end{align}

over a feasible set of $\sigma$. From the perspective of a windowed average, this roughly corresponds to choosing window length and weights that best fit the observed data. Figure \ref{fig:evidence_compare} demonstrates the influence of $\sigma$ on the smoothing posterior. When the scale is too small relative to the variation in the counts, the posterior intensity cannot change quickly enough to capture well the most recent reported counts and risks missing changes in trend; when the scale is increased, the posterior intensity can decrease quickly enough to explain the observations.

\section{Model monitoring and alerting} \label{sec:monitoring}

One of the UK government's main strategies for pandemic control has been to impose restrictions at the \gls{ltla} level in response to increasing case numbers. For our model to be useful in informing these decisions we must therefore have confidence in its predictions. Moreover, we would like to make assertions of probability on whether or not case numbers are following an increasing trend, or have exceeded a threshold of concern. We now describe how our model can be used to 1) monitor systematic reporting errors, and 2) act as an alert system at the \gls{ltla} level.

\subsection{Monitoring}
During the course of positive-test reporting thus far, there has been one major reporting error. A technical oversight resulted in the omission of 15,841 positive tests recorded between the $25^\text{th}$ of September and $2^\text{nd}$ of October, prior to a corrective backdate commenced on the $3^\text{rd}$ of October. This fundamental change in reporting should be detectable under a well-calibrated model. The posterior predictive distribution $\conditionaldist{x_{t+1}}{t}$ and lag-$j$ smoothing distribution $\conditionaldist{x_{t+1}}{t+j}$ each assign mass to a set of positive test counts at time $t+1$ which are credible according to the model, but accounting for observations up to times $t$ and $t+j$ respectively. 

When the reporting process is well matched with the assumptions of our model, we expect these distributions to be more consistent than when there are large systematic reporting errors. As a mechanism for detecting these errors, we propose the following consistency statistic:

\begin{align}\label{eq:x_consistency}
    C_t^{(j)} = \E_{\conditionaldist{x_{t+1}}{t+j}}\left[\conditionaldist{x_{t+1}}{t}\right].
\end{align}

When $j$ is large, the smoothing distribution reduces to a delta function on the truth, and the statistic amounts to measuring how well the predictive distribution captures the true value. Extending this reasoning to the sequence of lagged reports for time $t+1$, we recover the conditional evidence $\conditionaldist{\mathbf{y}_{t+1}}{t}$ by integrating out all unobserved random variables. Since $y_t^{(j)}$ under-reports $x_t$ when $j$ is small, we may choose to evaluate

\begin{align}\label{eq:ev_consistency}
    \conditionaldist{\mathbf{y}_{t-k:t+1}}{t-k-1} = \prod_{i=t-k}^{t+1} \conditionaldist{\mathbf{y}_{i+1}}{i}
\end{align}

as an aggregated measure of how well the model has captured reports which have not yet converged to a fixed value, which may constitute early warning signs against systematic reporting errors.

\subsection{Alerting} \label{sec:alerting}

\begin{figure}
    \centering
    \includegraphics[width=\textwidth]{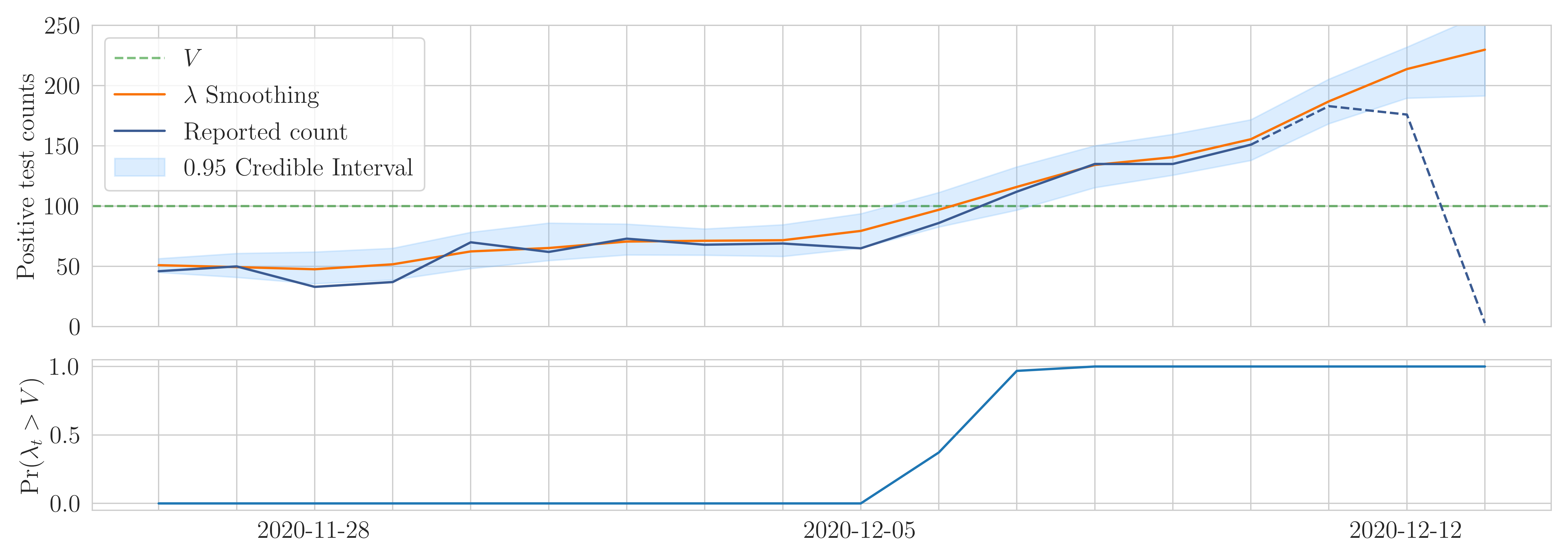}
    \caption{Example alert mechanism for Thurrock detecting if $\lambda$ is above a threshold value. \emph{Top}: the smoothing posterior on $\lambda$ as well as the reported counts and a threshold $V$. \emph{Bottom}: Probability that $\lambda$ is above the threshold value at each time step. } 
    \label{fig:thurrock_alert}
\end{figure}

\begin{figure}
    \centering
    \includegraphics[width=\textwidth]{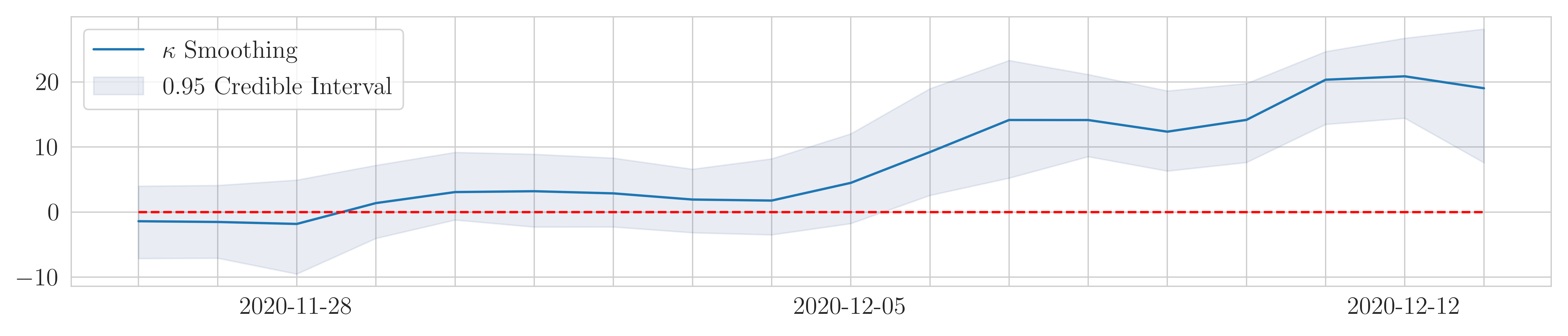}
    \caption{The smoothing posterior on the drift $\kappa_t$ for Thurrock showing $\lambda$ is following an increasing trend from $6^\text{th}$ December on-wards.} 
    \label{fig:thurrock_drift}
\end{figure}

A fundamental assumption of the model \eqref{eq:complete_joint} is that the positive test-counts $x_t$ at time $t$ are a-priori Poisson distributed with rate $\lambda_t$ on weekdays\footnote{On weekends, the rate is $\lambda_t z_t$ as discussed in Section \ref{sec:weekend_effect_model}}, and so $\text{Var}\left[x_t \mid \lambda_t \right] = \lambda_t$. Although windowed average techniques (see Section \ref{sec:kalman_filter}) may prove to be good estimators of the expected posterior on $\lambda_t$ for times $t$ such that we have observed $y_{t}^{(j)} = x_t$, these methods cannot include under-reports, or make statements of probability. For an alerting system that is both stable and probabilistic, we can examine the posterior behaviour of the $\lambda_{0:T}$ and $\kappa_{0:T}$. The difference $|\kappa_t - \kappa_{t+1}|$ may only substantially exceed $\sigma$ when there is sufficient posterior evidence to do so after observing $\mathbf{y}_{0:T}$. This makes the marginal smoothing posteriors $p(\lambda_t \mid \mathbf{y}_{0:T})$ and $p(\kappa_t \mid \mathbf{y}_{0:T})$ ideal for establishing if the intensity of the epidemic has crossed a threshold, and whether or not the sequence of intensities is increasing. For some threshold value $V$ we can use the fraction of smoothing particles which are larger than $V$

\begin{align}
    \conditionaldist{\lambda_t  > V}{T} &= \int_{V}^{\infty}\smoothingdist{\lambda_t}d\lambda_t\\
    &\approx \frac{1}{M}\sum_{j=1}^M \mathbbm{1}\left[\zeta_t^{(m)} > V\right]
\end{align}

to estimate an alerting probability. In Figure \ref{fig:thurrock_alert} we give an example of an alert for Thurrock between the $26^\text{th}$ of November and the $13^\text{th}$ of December and in Figure \ref{fig:thurrock_drift} we show how the smoothing posterior on $\kappa_{0:T}$ is easy to interpret visually as a description of whether the intensity is increasing.

\section{Results} \label{sec:nowcasting}

\begin{figure}
    \centering
    \includegraphics[width=\textwidth]{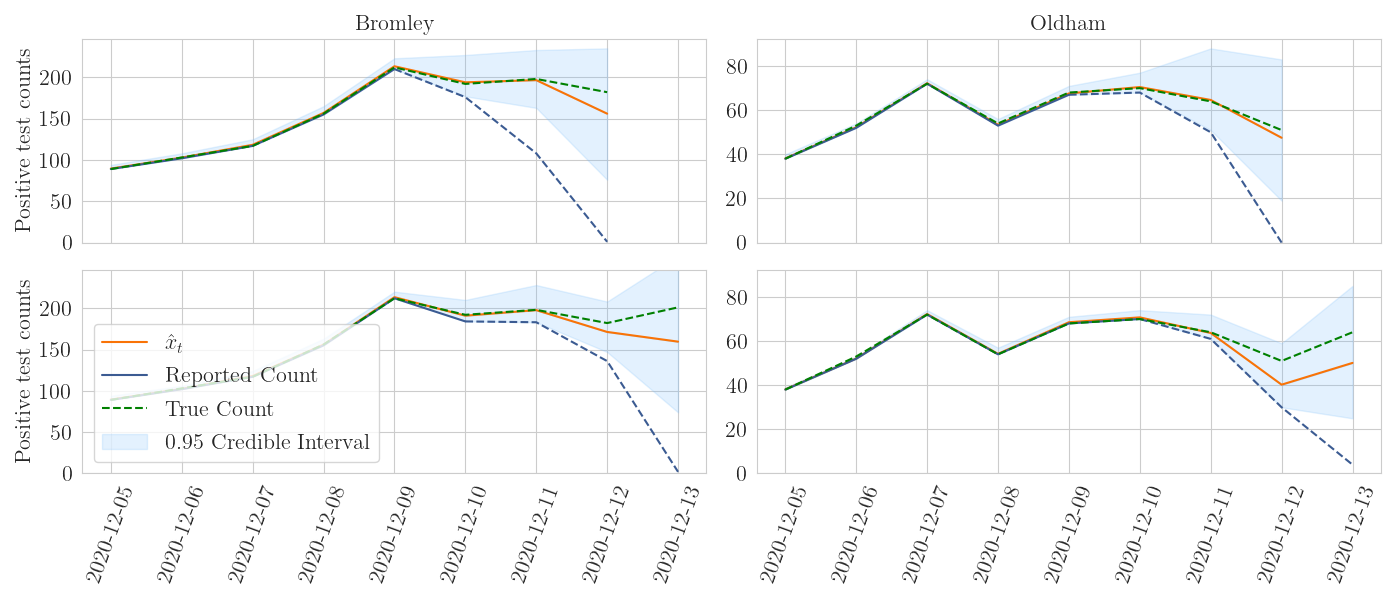}
    \caption{Example now-casts $p(x_t|y_{0:T})$ for Bromley and Oldham made on the 13\textsuperscript{th} of December and the 14\textsuperscript{th} of December. The posterior is exactly the true count for times long in the past. For recent times where we don't trust the reports, there is uncertainty as captured by the posterior distribution. As new and updated data is reported, all now-casts may be adjusted to account for this new information.}
    \label{fig:nowcast}
\end{figure}

Under the model described in this report, the now-cast on $x_T$ inferred from all available data is given by the smoothing distribution $\smoothingdist{x_T}$. In Figure \ref{fig:nowcast} we display this now-cast for two \glspl{ltla}, as well as the smoothing distribution for times $t < T$; the uncertainty on $x_t$ reduces with lag so for times far enough in the past the now-cast reflects our confidence that the reported count is the truth. We show results over a two week period in December when the UK was transitioning out of lockdown, where the model we propose may be the most useful in helping to monitor changes in the total number of positive tests. 

Though any statistical model on positive test cases provides the benefit of uncertainty quantification over non-statistical methods, the success of our model in deployment depends on its ability to improve upon simple techniques for estimating key indicators on the state of COVID-19. The 7-day windowed average on reports at the \gls{ltla} level, taken up to 4 days\footnote{This is the minimum omission period employed. When the reliability of recent counts has decreased, this window is extended.} before present day, is such a simple and publicly reported baseline. We may use the latest week of smoothing distributions to estimate the 7-day windowed average up to present day, rather than 4 days ago, so that decisions on local restrictions may incorporate information from under-reported count data. The models use a random walk scale selected to optimise the evidence as suggested by Section \ref{sec:model_selection}.

In absence of the full joint distribution $\smoothingdist{x_{t-7:T}}$ we estimate the lag-j 7-day windowed average for day $T$ $\text{WA}_T^{(j)}\left(7\right)$ from the marginals by

\begin{align}\label{eq:windowed_av}
    \text{WA}_T^{(j)}\left(7\right) = \frac{1}{7}\sum_{t=T-7}^T \E_{\conditionaldist{x_t}{T+j}}\left[ x_t \right]
\end{align}

To measure the performance of \eqref{eq:windowed_av} we assume the lag-7 reports have converged to $x_t$ and measure the absolute error

\begin{align}
    \text{AE}_T^{(j)}\left(7\right) = |\text{WA}_T^{(j)}\left(7\right) - \frac{1}{7}\sum_{t=T-7}^T y_t^{(7)}|
\end{align}

\begin{figure}
    \centering
    \includegraphics[width=\textwidth]{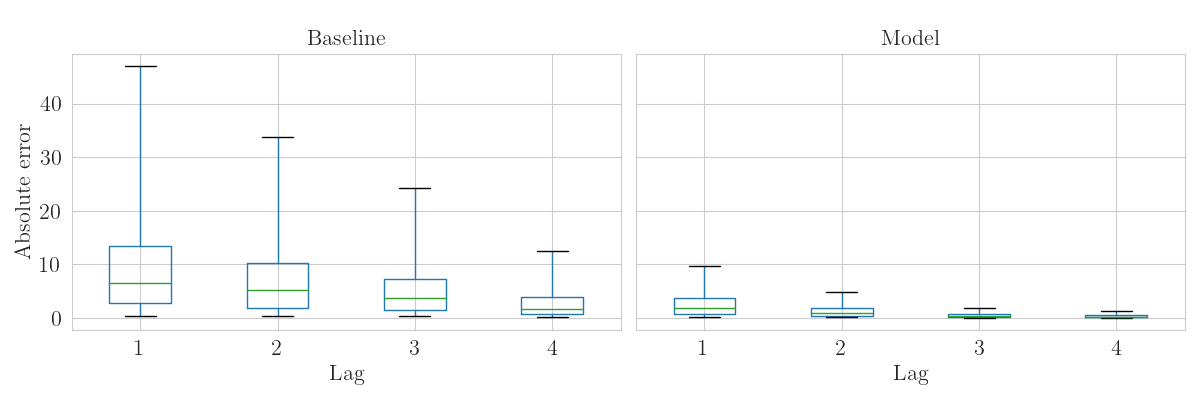}
    \caption{Distribution of absolute error for the imputed windowed average across all \glspl{ltla} by lag. The box plot whiskers cover the 5\textsuperscript{th} and 95\textsuperscript{th} percentiles. Model run on the 14\textsuperscript{th} of December reports.}
    \label{fig:ma_results}
\end{figure}

Figure \ref{fig:ma_results} shows the distribution of absolute errors for the imputed 7-day windowed average across all \glspl{ltla} alongside the performance of the current strategy of ignoring under-reports. It is clear that using the expected smoothing distribution as a now-cast improves the mean absolute error at early stages of reporting, and further that there is a significant reduction in the variance of this estimator compared to the baseline.

\section{Discussion}

We have presented a probabilistic model for time series count data with weekend effects and a lagged reporting process. We demonstrated that it is possible to use incomplete, lagged reports and impute the true count to obtain a seven day moving average that extends to the present date. Conditional on the now-cast being accurate, this is preferable to waiting for the most recent counts to converge before considering their use in monitoring the progress of the COVID-19 epidemic as it permits faster response to changes at the \gls{ltla} level. We also directly model the underlying intensity of the disease which lends itself naturally to probabilistic alerting strategies when incidence rates are increasing or have passed above some threshold value. 

Our approach relates closely to a number of concurrent works. In \cite{lancaster} a negative binomial model on swab tests is proposed, which is similar to that specified in Section \ref{sec:latent_process} when considering the marginal posterior on $x_{it}$ with $\lambda_{it}$ integrated out. They further posit a spatio-temporal dependence on the Poisson rates which extends the work described here. A key difference between \cite{lancaster} and here is the inference mechanism; \cite{lancaster} employs restricted maximum-likelihood parameter estimation, whilst we pursue a fully Bayesian treatment such that all parameter uncertainties are propagated through to quantities of interest. The model described in \cite{strathclyde} shares the same binomial likelihood as proposed here. They couple the 442 Scottish post-code districts spatially with a \gls{gmrf}. A number of internal notes from \gls{phe} have also studied outliers under binomial likelihood models with some spatio-temporal dependence \cite{phe1,phe2}. Generally, \glspl{ltla} which are spatially close may exhibit similar trends in the number of positive tests, and so we may expect a spatial coupling between neighbouring \glspl{ltla} to yield performance improvements. The most common models for dealing with spatial dependence are typically based on \glspl{gmrf}; the \gls{bym} model \cite{bym1991} posits spatial dependence through such a field, and although we do not employ this spatial structure here, a number of algorithms for posterior inference under this construction \cite{MORRIS2019100301, rue2009approximate} have been used to study COVID-19 \cite{dangelo2020spatial}. We note however that in this work we have observed mixed reliability in the spatial dependence of reporting rates.

Two recent works have also modelled a lagged reporting process \cite{Seaman2020.09.15.20194209, stoner2020}; both consider a temporal structure which codes calendar effects into the reporting lag, where the lagged counts are modelled as a generalised Dirichlet-multinomial distribution. This may deal more naturally with the reporting lag than our use of beta-binomial distributions on the latest available cumulative reports, though neither offers a posterior distribution on a latent intensity and would therefore have to rely on different strategies for probabilistic alerting than those justified in Section \ref{sec:alerting} here.

The accuracy of our model in now-casting the number of positive tests relies on stability of the reporting prior. We suggest that building a reporting prior against testing laboratories rather than \glspl{ltla} may make available tighter prior estimates for the reporting rate - since reports for an \gls{ltla} constitute an aggregation of reports from testing sites serving that authority, variance in the rates may arise from differences in reporting behaviour across these testing sites. In fact, since the end of October, each lab has provided capacity estimates in publicly available data\footnote{See \url{https://coronavirus.data.gov.uk/details/about-data}}. It may be possible therefore to estimate lag at the lab level, conditional on the testing load faced by each lab.

\section{Acknowledgements}

The authors thank the \gls{jbc} for the proposal of and input to this problem. In particular we acknowledge Dmitry Semashkov for significant engineering effort in the deployment of this model. We also thank Theo Sanderson for providing a consistent and up-to-date repository of snapshots on open data\footnote{\url{https://github.com/theosanderson/covid_uk_data_timestamped}}.

\clearpage
\begin{appendices}

\section{Marginal Filtering and Smoothing Derivations} \label{app:derivations}

 In what follows we specify filtering and smoothing distributions for each time step $t$, from which we can draw samples but cannot compute in closed form. Denoting these samples by $\Lambda_t^{(i)}, K_t^{(i)} \sim p(\lambda_t, \kappa_t \mid y_t, \ldots, y_0)$ and $L_t^{(j)},\zeta_t^{(j)} \sim p(\lambda_t, \kappa_t \mid y_T, \ldots, y_0)$, we build the atomic measures

\begin{align}
 &\filteringdist{\lambda_t, \kappa_t} \approx \frac{1}{N}\sum_{i=1}^N \deltafunc{\lambda_t - \Lambda_t^{(i)}}\deltafunc{\kappa_t - K_t^{(i)}}\\
 &\smoothingdist{\lambda_t, \kappa_t} \approx \frac{1}{M}\sum_{j=1}^M \deltafunc{\lambda_t - L_t^{(j)}}\deltafunc{\kappa_t - \zeta_t^{(j)}}
\end{align}

as particle approximations whose accuracy is determined by the number of samples $N$ and $M$ and the nature of the sampling scheme employed to draw these samples.

\subsection{Filtering} \label{sec:appendix_filtering}
For the forward pass we build a sequence of conditional distributions by

\begin{align}
\predictivedist{\lambda_t, \kappa_t, \mathbf{y}_t} &= \int \predictivedist{\lambda_t, \lambda_{t-1}, \kappa_t, \kappa_{t-1}, \mathbf{y}_t} d\lambda_{t-1}d\kappa_{t-1}\\
&=  p(y_t \mid \lambda_t)\int p(\lambda_t, \kappa_t \mid \lambda_{t-1}, \kappa_{t-1})p(\lambda_{t-1}, \kappa_{t-1} \mid y_{0:t-1})d\kappa_{t-1} d\lambda_{t-1}\\
&= p(y_t \mid \lambda_t) \int  p(\lambda_{t} \mid \kappa_t, \lambda_{t-1})p(\kappa_t \mid \kappa_{t-1})p(\lambda_{t-1}, \kappa_{t-1} \mid y_{0:t-1})d\kappa_{t-1} d\lambda_{t-1}\\
&\approx \frac{p(y_t \mid \lambda_t)}{N}\sum_{i=1}^{N} \delta\left(\lambda_t - (\kappa_t + \Lambda_{t-1}^{(i)})\right)\mathcal{N}\left(\kappa_t; K_{t-1}^{(i)}, \sigma^2\right)
\end{align}

from which we can draw samples with \gls{mh} by first proposing a step in $\kappa_t$ from a symmetric distribution, and then drawing $\lambda_t$ uniformly from $\{\kappa_t + \Lambda_{t-1}^{(1)}, \ldots, \kappa_t+ \Lambda_{t-1}^{(N)}\}$. 

\subsection{Smoothing} \label{sec:appendix_smoothing}
We may use the conditional dependency structure of the model to write

\begin{align}
\smoothingdist{\lambda_t, \kappa_t} &= \int \smoothingdist{\lambda_t, \kappa_t, \lambda_{t+1}, \kappa_{t+1}} d\lambda_{t+1}d\kappa_{t+1}\\
&= \int p(\lambda_t, \kappa_{t} \mid \lambda_{t+1}, \kappa_{t+1} \mathbf{y}_{0:t})\smoothingdist{\lambda_{t+1}, \kappa_{t+1}}d\lambda_{t+1}d\kappa_{t+1}\\
&= \filteringdist{\lambda_t, \kappa_t}\int \frac{p(\lambda_{t+1}, \kappa_{t+1} \mid \lambda_{t}, \kappa_{t})}{\filteringdist{\lambda_{t+1}, \kappa_{t+1}}}\smoothingdist{\lambda_{t+1}, \kappa_{t+1}}d\lambda_{t+1}d\kappa_{t+1}\\
&= \filteringdist{\lambda_t, \kappa_t}\int \frac{p(\lambda_{t+1} \mid \kappa_{t+1}, \lambda_{t})p(\kappa_{t+1} \mid \kappa_t)}{\filteringdist{\lambda_{t+1}, \kappa_{t+1}}}\smoothingdist{\lambda_{t+1}, \kappa_{t+1}}d\lambda_{t+1}d\kappa_{t+1}\\
&\approx \frac{\filteringdist{\lambda_t, \kappa_t}}{M} \sum_{j=1}^{M} \frac{p(L_{t+1}^{(j)} \mid \zeta_{t+1}^{(j)}, \lambda_t)p(\zeta_{t+1}^{(j)} \mid \kappa_t)}{\filteringdist{L_{t+1}^{(j)}, \zeta_{t+1}^{(j)}}}\label{eq:full_smoothing}
\end{align}

Let $w_{ij} = \deltafunc{L_{t+1}^{(j)} - (\zeta_{t+1}^{(j)} + \Lambda_t^{(i)})}\normaldist{\zeta_{t+1}^{(j)}; K_t^{(i)}}{\sigma^2}$, $w_{*j}=\sum_{i=1}^Nw_{ij}$. Replacing the filtering distribution in \eqref{eq:full_smoothing} by its atomic measure we obtain


\begin{align}
  \smoothingdist{\lambda_t = \Lambda_{t}^{(i)}, \kappa_t = K_{t}^{(i)}} \approx \frac{1}{M}\sum_{j=1}^M \frac{w_{ij}}{w_{*j}}\label{eq:smoothing_weights}
\end{align}

which indicates that each smoothing step is a multinomial resampling of the corresponding filtering atoms with weights \eqref{eq:smoothing_weights}. For smoothing the counts $x_t$ we begin with the joint




\begin{align}
    p(\lambda_t, x_t \mid \mathbf{y}_{0:T}) &= p(x_t \mid \lambda_t, \mathbf{y}_{0:T}) p(\lambda_t \mid \mathbf{y}_{0:T})\\
    &= p(x_t \mid \lambda_t, \mathbf{y}_t)p(\lambda_t \mid \mathbf{y}_{0:T})\\
\end{align}

and so

\begin{align}
    \smoothingdist{x_t} & = \int \smoothingdist{x_t, \lambda_t} d\lambda_t\\
    &\approx \frac{1}{M} \sum_{j=1}^M p(x_t \mid \zeta_t^{(j)}, \mathbf{y}_t)\\
    &\approx \frac{p(\mathbf{y}_t \mid x_t)}{M} \sum_{j=1}^M \frac{p(x_t \mid \zeta_{t}^{(j)})}{p(\mathbf{y}_t \mid \zeta_t^{(j)})} \label{eq:x_smoother}
\end{align}

so that if we have the smoothing atoms $\{\zeta_t^{(j)}\}_{t=1,j=1}^{T, M}$ then we can evaluate \eqref{eq:x_smoother}. With the addition of weekend effects we must alter the smoothing procedure for each $x_t$ by considering

\begin{align}
    \smoothingdist{x_t, \lambda_t, z_t} &= p(x_t \mid \lambda_t, z_t, \mathbf{y}_{0:T})\smoothingdist{\lambda_t, z_t}\\
    & = p(x_t \mid \lambda_t, z_t, \mathbf{y}_t)\smoothingdist{\lambda_t, z_t}\\
    & = \frac{p(\mathbf{y}_t \mid x_t)p(x_t \mid \lambda_t, z_t)}{p(\mathbf{y}_t \mid \lambda_t, z_t)} \smoothingdist{\lambda_t, z_t} \label{eq:weekend_joint_smooth}
\end{align}

Now for the joint smoothing distribution on $\lambda_t, z_t$ we have

\begin{align}
 \smoothingdist{\lambda_t, z_t} &= p(z_t \mid \lambda_t, \mathbf{y}_{0:T})\smoothingdist{\lambda_t}\\
 &= p(z_t \mid \lambda_t, \mathbf{y}_t)\smoothingdist{\lambda_t}\\
 &= \frac{p(\mathbf{y}_t, z_t \mid \lambda_t)}{p(\mathbf{y}_t \mid \lambda_t)}\smoothingdist{\lambda_t}\\
 &= \frac{p(\mathbf{y}_t \mid \lambda_t, z_t)p(z_t)}{p(\mathbf{y}_t \mid \lambda_t)}\smoothingdist{\lambda_t}
\end{align}

which we may employ to evaluate \eqref{eq:weekend_joint_smooth}.

\section{Evidence}
By the sum and product rules of probability we may write

\begin{align}
  p(\mathbf{y}_{1:T}) &= p(\mathbf{y}_0)\prod_{t=1}^{T} \predictivedist{\mathbf{y}_t}\\
  &= p(\mathbf{y}_0)\prod_{t=1}^{T} \int \predictivedist{\mathbf{y}_t, \lambda_t, \lambda_{t-1}, \kappa_t, \kappa_{t-1}} d\lambda_t d\lambda_{t-1} d\kappa_t d\kappa_{t-1}\\
  &= p(\mathbf{y}_0)\prod_{t=1}^{T} \int p(\mathbf{y}_t \mid \lambda_t) p(\lambda_t \mid \lambda_{t-1}, \kappa_t) \predictivedist{\lambda_{t-1}, \kappa_{t-1}}p(\kappa_t \mid \kappa_{t-1})d\lambda_t d\lambda_{t-1}\\
  &\approx p(\mathbf{y}_0)\prod_{t=1}^{T} \frac{1}{N}\int p(\mathbf{y}_t \mid \lambda_t) \sum_{i=1}^{N} \deltafunc{\lambda_t - (\Lambda_{t-1}^{(i)} + \kappa_t)} \normaldist{\kappa_t; K_{t-1}^{(i)}}{\sigma^2} d\lambda_{t} \label{eq:evidence_mixture}
\end{align}

in \eqref{eq:evidence_mixture}, we can also approximate the remaining integral by sampling from
the Gaussian mixture built from transition distributions centered on the filtering particles at the previous time step. This yields an approximation to the evidence at each time step; sampling some $\kappa_t = K_t^{(i)}$ immediately implies which $\lambda_t = \Lambda_{t-1}^{(i)} + K_t^{(i)}$.









\section{The most recent count $\itlast{y}$ is sufficient for $x_{it}$}
\label{appendix:suff_stat_x}
Defining $z_{it}^{(\cdot)}$ and $\phi_{it}^{(\cdot)}$ to be the first-order differences of $y_{it}^{(\cdot)}$ and $\theta_{it}^{(\cdot)}$ respectively, i.e.\ for any $J \in \{1,\ldots, T-t\}$ we have
\begin{align}
\label{eq:yzdef}
    y_{it}^{(J)} &\equiv \sum_{j=1}^J z_{it}^{(j)}\\
\label{eq:thetaphidef}
    \theta_{it}^{(J)} &\equiv \sum_{j=1}^J \phi_{it}^{(j)}\ ,
\end{align}
and under a multinomial model for $z_{it}^{(1:J)}$, 
\begin{align}
\label{eq:multinomlik}
    z_{it}^{(1:J)}\mid x_{it}, \phi_{it}^{(1:J)} &\sim  \mathrm{Multinomial}(x_{it}, \phi_{it}^{(1:J)})\ ,
\end{align}
we demonstrate here that $y_{it}^{(J)}$ is a sufficient statistic for $x_{it}$, i.e.\ that
\begin{align*}
    p(y_{it}^{(1:J)} \mid x_{it}, \theta_{it}^{(1:J)}) & \overset{x_{it}}{\propto} p(y_{it}^{(J)} \mid x_{it}, \theta_{it}^{(J)})
\end{align*}
where $\overset{x_{it}}{\propto}$ denotes proportionality with respect to $x_{it}$. 

It is clear by definition \eqref{eq:yzdef} that either of $z_{it}^{(1:J)}$ and $y_{it}^{(1:J)}$ can be reconstructed from the other; they therefore contain equivalent information on $x_{it}$. The analogous joint reconstruction property in \eqref{eq:thetaphidef} allows us to condition interchangeably on $\theta_{it}^{(1:J)}$ and $\phi_{it}^{(1:J)}$.
Sufficiency of $y_{it}^{(J)}$ for $x_{it}$ then becomes clear when viewed through the multinomial likelihood at \eqref{eq:multinomlik}:
\begin{align*}
    p(y_{it}^{(1:J)} \mid x_{it}, \theta_{it}^{(1:J)}) & \overset{x_{it}}{\propto} p(z_{it}^{(1:J)} \mid x_{it}, \phi_{it}^{(1:J)})\\
    & = \frac{x_{it}!}{(x_{it} - \sum_{j=1}^J z_{it}^{(j)})! \prod_{j=1}^J z_{it}^{(j)}!} \left(1 - \sum_{j=1}^J\phi_{it}^{(j)}\right)^{x_{it} - \sum_{j=1}^J z_{it}^{(j)}} \prod_{j=1}^J (\phi_{it}^{(j)})^{z_{it}^{(j)}}\\
    & \overset{x_{it}}{\propto} \frac{x_{it}!}{(x_{it} - y_{it}^{(J)})!} \left(1 - \theta_{it}^{(J)}\right)^{x_{it}  - y_{it}^{(J)}}\\
    & \overset{x_{it}}{\propto}  \mathrm{Binomial}(y_{it}^{(J)} \mid x_{it}, \theta_{it}^{(J)})\ .
\end{align*}


\section{Kalman Gain} 
\label{sec:kalman_gain}

Let $\text{v}_t = \text{Var}_{p(x_t \mid y_{0:t})}(x_t)$ be the variance of the filtering posterior at time-step $t$. The Kalman Gain is given by

\begin{align}
    K_t = \frac{\sigma^2 + \text{v}_{t-1}}{\sigma^2 + \text{v}_{t-1} + \sigma_y^2}    
\end{align}

which in effect controls the weight given to the most recent data point in computing the filtering posterior; the relative scale of $\text{v}_{t-1}$ and $\sigma_y$ determines this weighting.

\end{appendices}

\bibliographystyle{unsrt}
\bibliography{refs}

\end{document}